\documentclass[traditabstract]{aa} 
\usepackage{graphicx}
\usepackage{txfonts}
\usepackage{natbib}
\bibpunct{(}{)}{;}{a}{}{,}

\newcommand{\ux}{UX~Mon }
\newcommand{\uxe}{UX~Mon}

\newcommand{\ubv}{\hbox{$U\!B{}V$}}

\newcommand{\oc}{\hbox{$O\!-\!C$}}

\newcommand{\deri}{\rm d}

\newcommand{\kms}{km~s$^{-1}$ }
\newcommand{\ks}{km~s$^{-1}$}

\newcommand{\ms}{M$_{\odot}$}

\newcommand{\ha}{H$\alpha$ }

\begin{document}

   \title{UX Mon as a W Ser binary}

   \author{D. Sudar
          \inst{1}
          \and
          P. Harmanec
          \inst{2}
          \and
          H. Lehmann
          \inst{3}
          \and
          S. Yang
          \inst{4}
          \and
          H. Bo\v{z}i\'{c}
          \inst{1}
          \and
          D. Ru\v{z}djak
          \inst{1}
          }

   \institute{Hvar Observatory, Faculty of Geodesy,
              Ka\v{c}i\'{c}eva 26, University of Zagreb, 10000 Zagreb, Croatia
         \and
             Astronomical Institute of the Charles University, Faculty of Mathematics and Physics,\\
             V Hole\v{s}ovi\v{c}k\'{a}ch 2, CZ-180 00 Praha 8, Czech Republic
         \and
             Th\"{u}ringer Landessternwarte Tautenburg, 07778 Tautenburg, Germany
         \and
             Dept. of Physics and Astronomy, University of Victoria, PO Box 3055, Victoria BC V8W 3P6, Canada
             }

\offprints{D. Sudar, \email: dsudar@geof.hr}

   \date{Release \today}

\abstract{Using our new photometric and spectroscopic observations as well as
all available published data, we present a new interpretation of
the properties of the peculiar emission-line binary UX Mon. We conclude that
this binary is in a rare phase of fast mass transfer between the binary
components prior to the mass ratio reversal. We firmly establish that
the orbital period is secularly decreasing at a rate of
$\dot{P}=-0.260$ seconds per year. From several lines of reasoning,
we show that the mass ratio of the component losing mass
to the mass-gaining component $q$ must be larger than 1 and find our
most probable value to be $q=1.15\pm0.1$.
The BINSYN suite of programs and the steepest descent method
were used to perform the final modeling. We modeled the star as a~W~Ser star
with a thick disk around its primary. Although the remaining
uncertainties in some of the basic physical elements describing the system
in our model are not negligible, the model is
in fair agreement with available observations.
Only the nature of the light variations outside
the primary eclipse remains unexplained.

}

   \maketitle

\section{Introduction}

\object{UX Monocerotis} (HD 65607, SAO 135333) is an eclipsing binary with an
orbital period of 5\fd90. Its binary nature was first
reported by \cite{Woods1928}, who also noted that its brightness at maximum
light is not constant. \citet{Gaposchkin1947} studied 32 spectrograms from the
McDonald Observatory and 1389 photographic plates from the Harvard
Observatory. He identified the star eclipsed at the deeper minimum
(hereafter, the primary) as an A7p object and the other component (hereafter,
the secondary) as a G2p star. He arrived at surprisingly low masses of
$M_{1}=0.73$~\ms\ and $M_{2}=0.74$~\ms\ for the primary
and the secondary, respectively. \citet{Struve1947} analyzed 152
photographic spectra secured in 1947. He obtained radial velocity
(RV hereafter) semi-amplitudes of $K_{1}=140$~\kms and $K_{2}=60$~\kms for
the primary and secondary, respectively. This implies a more massive
secondary, $M_{2}\sin^{3}i=3.4$~\ms, than the primary,
$M_{1}\sin^{3}i=1.5$~\ms, which would mean that \ux is
in the rare evolutionary stage prior to the mass reversal
\citep[cf. e.g.][]{Crawford1955,Morton1960}. Struve also discussed
the origin of the Balmer emission, which he identified as circumstellar
matter located between the two stars.

\citet{Hiltner1950} published 486 yellow photoelectric observations with
an effective wavelength of 5300~\AA . They confirmed the light variations
outside the primary eclipse. They also noted that
the durations of both eclipses are the same. \citet{Wood1957} obtained
light curves of the system in the {\it yellow} and {\it blue}
filters. He observed short (hours) and long (six or more
days) non-orbital light changes. \citet{Lynds1956,Lynds1957a,Lynds1957b}
secured and analyzed 333 \ubv\ observations. He
derived the ratio of radii, $k=0.522$, the radius of the secondary relative to the orbital separation,
$r_{2}=0.358$, and the inclination, $i=83\fdg 6$.

\citet{Scaltriti1973, Scaltriti1976} concluded that the primary is a $\delta$
Scuti variable and that there had been an abrupt change of the orbital period in the
past.

\cite{Umana1991} found that the UX Mon system is a microwave source
and concluded that the microwave radiation originates from the gyrosynchrotron
radiation of electrons in a magnetic field.

\citet{Olson1995} carried out a detailed photometric and spectroscopic
study of \ux based on their new $ubvyI$ observations and a series of
electronic spectra. They derived the most accurate RV semi-amplitude
of the secondary yet obtained, $K_{2}=$108.3$\pm$1.9 km s$^{-1}$, but were unable to detect
any measurable spectral lines of the primary. In contrast to \citet{Struve1947},
they inferred a mass ratio secondary/primary of $q=0.8$, based on light
curves analysis. Their model also led to the surface temperatures
$T_{1}\approx$ 8000 K and $T_{2}\approx$ 5500 K for the primary and secondary,
respectively. They concluded that the gas stream from the secondary
hits the primary directly, without forming an accretion disk around it.

Analyzing published photometry and spectroscopy, \citet{Ondrich2003}
concluded that the orbital period of \ux has been secularly decreasing and
provided new quadratic ephemeris. They also discussed the emission
lines of highly ionized atoms in the HST/GHRS spectra indicating that \ux
is a W~Ser star.
W~Ser stars are strongly interacting binaries characterized by
the presence of numerous emission lines in their UV spectra, which are believed
to originate in an extended gaseous envelope around their mass-gaining
components \citep[cf. e.g.][]{plavec78, Plavec1980, Plavec1992}.
However, the mass transfer rate ($10^{-8}-10^{-9}$ \ms\ y$^{-1}$),
which \citet{Ondrich2003} derived for \ux from
the quadratic ephemeris, appears to be too low compared to the typical values of
$10^{-6}$ \ms\ y$^{-1}$ measured for W~Ser binaries
\citep{Plavec1980, Plavec1992}. Independently, the secular period
decrease was also noted by \citet{Meyer2006}.

\citet{Olson2009}
published a new study of \ux based on systematic photometry secured
over a period of eight years.
They found almost the same system parameters as in their previous paper \citep{Olson1995}.
Focusing mainly on light variations outside the primary eclipse, they
concluded that these are probably caused by the variable mass transfer.

\section{Observations}
\subsection{Photometry}

\begin{table}
\caption{Journal of photometric observations of \uxe}
\label{PhotData}
\begin{center}
\begin{tabular}{c c c c}
\hline\hline
Sources & Time interval & No. of. & Passbands \\
 & (HJD-2400000) & obs &  \\
\hline
 (1) & 18600.0-18605.7 & 38 & vis \\
 (2) & 33246.9-33381.6 & 486 & $\lambda$5300 \\
 (3) & 33300.8-33399.7 & 401 & $\lambda$5200, $\lambda$4250 \\
 (4) & 35084.9-35153.9 & 999 & $UBV$ \\
 (5) & 41626.7-41764.3 & 875 & $\lambda$5150, V \\
 (6) & 47181.7-48656.9 & 5056 & $uvbyI$ \\
 (7) & 47964.6-48938.2 & 76 & $V$ \\
 (8) & 51868.8-53896.5 & 268 & $V$ \\
 (9) & 52655.4-54219.3 & 402 & $UBV$ \\
 (10) & 52655.4-54219.3 & 137 & vis \\
 (11) & 51448.0-54228.6 & 1453 & $BV$ \\
\hline
\end{tabular}
\end{center}
References. (1) \citet{Woods1928}; (2) \citet{Hiltner1950};
(3) \citet{Wood1957}; (4) \citet{Lynds1957a}; (5) \citet{Scaltriti1973};
(6) \citet{Olson1995}; (7) \citet{Perryman1997}; (8) \citealt{Pojmanski1997};
(9) Hvar observations (this paper); (10) \citet{Meyer2007}; (11) \citet{Olson2009}.
\end{table}

Altogether, 22 different datasets of photometric observations of \ux
from 11 different sources were used. Table~\ref{PhotData} lists basic observational data.
More information about the datasets used in this paper is given below.
\begin{itemize}
\item[(1)] Photographic photometry, consisting of 38
phase-averaged normal points published by \cite{Woods1928} only
in orbital phases. The accurate dates of observations are unknown
limiting their usefulness in our study.

\item[(2)] These observations were obtained at McDonald Observatory
from November 26, 1949 to April 10, 1950.

\item[(3)] \cite{Wood1957} obtained
401 photoelectric observations, divided into two ({\it green and blue})
datasets.

\item[(4)] There are 333 individual
observations in each of the \ubv\ passbands in this dataset, obtained during twenty nights
between December 7, 1954, and February 15, 1955. The instrumental
color system is essentially the same as the standard $UBV$ system, except
for the zero point.

\item[(5)] Two datasets of green and yellow photoelectric observations
contain a total of 875 individual observations.

\item[(6)] These five sets of $uvbyI(Kron)$ observations consist of 5056
observations in total. They were secured between January 21, 1988, and
February 4, 1992 and analyzed by \cite{Olson1995}. Upon our request, Dr.~Olson
kindly allowed us to have copies of these observations at our disposal. These data were
subsequently published by \citet{Olson2009}.

\item[(7)] The Hipparcos $H_{p}$ photometry of UX Monocerotis was extracted
from the data archive published by \cite{Perryman1997} and transformed
to the standard Johnson $V$ magnitude using the transformation
formulae derived by \citet{Harmanec1998} and \citet{Harmanec2001},
which are based on numerous all-sky \ubv\ observations of many stars
observed at Hvar.

\item[(8)] All-Sky Automated Survey (ASAS hereafter) is an observational
program for monitoring stars of the southern hemisphere brighter than
 $\approx$14th magnitude \citep{Pojmanski1997, Pojmanski2001}.
Altogether, 268 Johnson $V$ observations of \ux were secured
between November 20, 2000 and June 10, 2006.

\item[(9)] Altogether, 402 new photoelectric \ubv\ observations were secured
during two seasons with the 0.65-m Cassegrain reflector of the Hvar Observatory.
The transformation from the instrumental to the standard Johnson system was
carried out by the program HEC22 \citep{Harmanec1994}, which uses non-linear
transformation formulae.  The most developed version rel.16.2 was used,
which allows modeling of variable extinction during the observing
nights.

\item[(10)] This dataset consists of visual estimates
by \cite{Meyer2006,Meyer2007}.

\item[(11)] \cite{Olson2009} obtained 1453 individual measurements in
Johnson's $B$ and $V$ filters with the 0.4m APT at Fairborn
Observatory in southern Arizona, USA.

\end{itemize}

\subsection{Spectroscopy}
\label{SectSpect}

In total, 39 new electronic spectra obtained at three observatories are included in this paper.
\begin{itemize}
\item[(1)]Six spectra were obtained at the coud\'e focus of the Ond\v{r}ejov
Observatory 2.0-m reflector with a linear dispersion of 17~\AA\ mm$^{-1}$ (two-pixel
resolution 12600), which cover a spectral range 6300-6700~\AA.  Subtraction of bias, flatfielding,
creation of 1-D spectra, and wavelength calibration has been routinely carried out
shortly after the observations by Dr.~M.~\v{S}lechta.
Rectification, heliocentric correction of time and
radial velocity, and the removal of cosmic spikes was carried out by DS
with the help of the program SPEFO \citep{Horn1996,Skoda1996} in its form
developed by the late Mr.~J.~Krpata.
\item[(2)]Thirteen spectra were obtained in the coud\'e focus of the
1.22-m reflector of the Dominion Astrophysical Observatory (DAO hereafter).
Their linear dispersion is 10~\AA\,mm$^{-1}$ (two-pixel
resolution 21700), while the range of wavelengths
is 6100-6700~\AA. The initial reduction of spectra (bias removal, flatfielding,
and creation of 1-D spectra) was carried out by SY using IRAF.
The wavelength calibration and all subsequent reductions were carried out
by DS in SPEFO.
\item[(3)]Twenty echelle spectra were obtained at the coud\'e focus of the
2.0-m reflector of TLS Tautenburg. These spectra have the best linear
dispersion of all three sets: 3.2~\AA\ mm$^{-1}$ (two-pixel
resolution 63000). They cover
the spectral range 4700-7085~\AA.
All reductions, including the rectification and heliocentric
corrections were carried out by HL using standard MIDAS packages
and a special routine to determine the nightly instrumental RV zero
points from a large number of telluric O$_{2}$ lines in the spectra.
\end{itemize}

\section{Analysis}

\subsection{Secular changes of the orbital period}

\cite{Ondrich2003} proposed that there had been a secular {\sl decrease}
of the orbital period. We decided to re-investigate their finding, using
more photometric observations than they had at their disposal.

We first calculated the local linear ephemeris of the sources listed in Table~\ref{PhotData} with a
sufficient number of measurements by using the latest publicly available version
of the program FOTEL \citep{Hadrava2004b}. We allowed for the convergence of the epoch of the primary minimum
and the local period.  The results are summarized
in Table \ref{tab_lin_eph}. We also added a third column, \oc$_2$, listing
differences between the local epochs
and those calculated from the {\sl quadratic ephemeris} (see below).

\begin{table}
\caption{Times of the primary minima and values of the instantaneous
orbital period as derived by FOTEL assuming locally linear ephemeris.
The column \oc$_2$ gives the differences with respect
to {\sl quadratic} ephemeris (Eq. \ref{Eq_quad_ephem}).}
\label{tab_lin_eph}
\begin{center}
\begin{tabular}{c c c r}
\hline\hline
Source & Epoch          & Period & \oc$_2$\\
       & (HJD-2400000)  & (days) & (days)\\
\hline
(2) & 33246.1720(092) & 5.905882(563) & -0.0053\\
(3) & 33299.3269(063) & 5.904718(520) & +0.0082\\
(4) & 35088.3882(047) & 5.904210(689) & -0.0208\\
(5) & 41630.6478(157) & 5.904829(920) & -0.0074\\
(6) & 47180.8866(007) & 5.904494(004) & -0.0028\\
(8) & 51869.0342(078) & 5.904404(038) & +0.0016\\
(9) & 53746.6472(038) & 5.904432(067) & +0.0042\\
(11) & 52772.4039(019) & 5.904378(013) & -0.0008\\
\hline
\end{tabular}
\end{center}
\end{table}

In FOTEL, there is also an option that allows us to derive a period change,
$\dot P$, that exactly corresponds to a quadratic ephemeris, as one of the
unknowns of the solution. As discussed by \citet{Harmanec1993},
a quadratic ephemeris given by
\begin{equation}
T=T_0+P_0\cdot E + a\cdot E^2,\label{quadr}
\end{equation}
\noindent holds for the times of minima, where $P_0$ denotes
the value of the period at the time of reference minimum $T_0$.
Treating the epoch $E$ as a {\sl real number} characterizing
the cycle and phase, and denoting $T$ as the time corresponding to $E$
(i.e. to the given orbital phase at that cycle), one can write the
instantaneous orbital period, $P$, as
\begin{equation}
P={\deri T\over{\deri E}}=P_0+2aE\,.\label{Pted}
\end{equation}
\noindent The instantaneous change of the period, $\dot P$, is then
\begin{eqnarray}
\dot P=2a{\deri E\over{\deri T}}={2a\over{P}}\,.\label{dP}
\end{eqnarray}
This shows that the quadratic ephemeris leads to a linear change
of the period with epoch {\sl but not with time}. For instance, for
negative values of $a$, the rate of the period change
{\sl increases with time}.

We therefore used the corresponding option in FOTEL to derive the quadratic
ephemeris given in Table~\ref{FOTELeph} by combining all datasets from
Table~\ref{tab_lin_eph}.

\begin{table}
\caption{The quadratic ephemeris, based on all photometric datasets
of Table~~\ref{tab_lin_eph}, derived with FOTEL.}
\label{FOTELeph}
\begin{center}
\begin{tabular}{c c}
\hline\hline
Quantity & Value\\
\hline
$T_{\rm prim.min.}$ [HJD]& $2452666.1319\pm0.0012$\\
$P_{0}$ [days]& $5.9044365\pm0.0000020$\\
$\dot{P}_{T=T_{0}}$ [days per day]& $(-8.23\pm0.18)\times 10^{-9}$\\
$a$ [days]& $(-2.43\pm0.05)\times 10^{-8}$\\
\hline
\end{tabular}
\end{center}
\end{table}

The explicit expression for the quadratic ephemeris is
\begin{equation}
T_{\rm min.I}={\rm HJD} 2452666.1319+5.9044365E-2.43\times10^{-8}E^{2},
\label{Eq_quad_ephem}
\end{equation}
where the coefficient of the quadratic term is calculated as $P_{0}\dot{P}_{T=T_{0}}/2$.
The change of the orbital period corresponds to $\dot{P}_{T=T_{0}}=-0.260$ seconds per year.

\begin{figure}
\resizebox{\hsize}{!}{\includegraphics{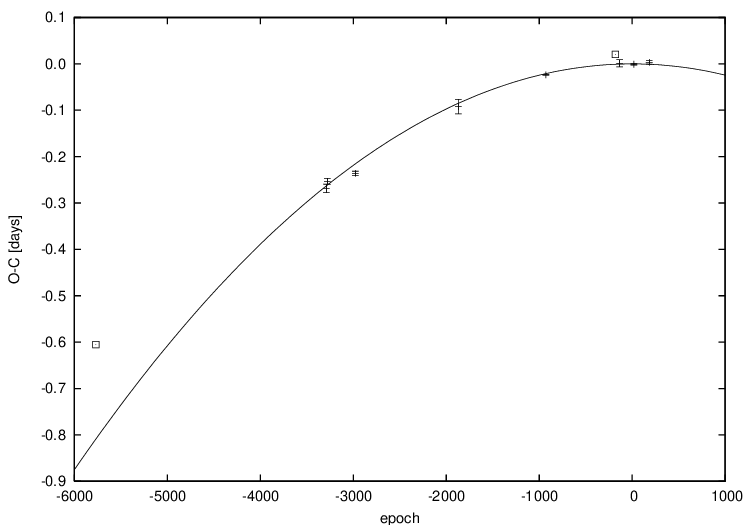}}
\caption{\oc\ in days with respect to the linear ephemeris. Crosses with
error bars denote the times of minima from Table~\ref{tab_lin_eph}.
The empty squares denote epochs of minima of sources (1) and (10).
The times of individual observations for source (1) are unknown,
and measurements from source (10) are of insufficient accuracy. Quadratic ephemeris
is shown with a solid curve.}
\label{ephem2}
\end{figure}

The local epochs of minima from Table~\ref{tab_lin_eph}
were compared to the global linear ephemeris and
the corresponding \oc\ deviations are shown in Fig.~\ref{ephem2}.
They clearly show the parabolic change indicating a secular
{\sl decrease} of the orbital period, in accordance with the findings
of \citet{Ondrich2003}.

Note that the epoch of the early photographic observations by \citet{Woods1928}
does not fit well with the extrapolated prediction of the quadratic ephemeris
(Eq.~\ref{Eq_quad_ephem}), but it also confirms that the period has been
decreasing. Since the dates of these early observations
are not known, we could not include this dataset into our FOTEL solution
to obtain yet more accurate values for the quadratic ephemeris and the rate of the period
change.

\begin{figure}
\resizebox{\hsize}{!}{\includegraphics{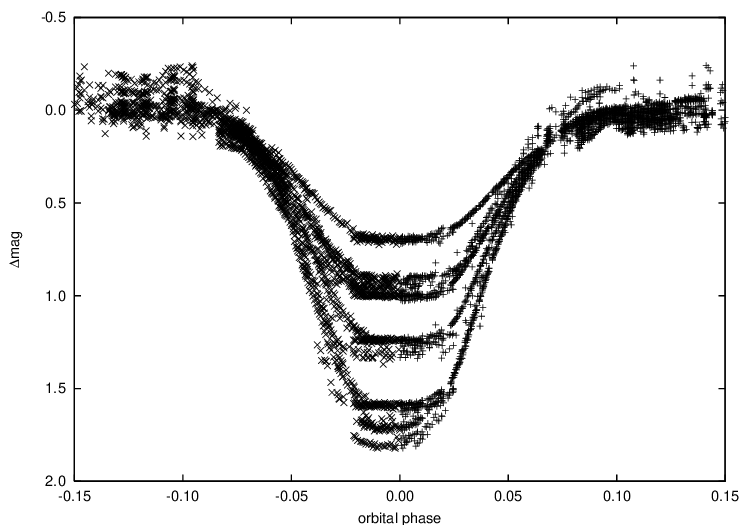}}
\caption{A phase plot of all photometric observations around the phase
of primary minimum using quadratic ephemeris (Eq. \ref{Eq_quad_ephem}).}
\label{corr_eph}
\end{figure}

In Fig.~\ref{corr_eph} we fold all photometry near the primary minima from the datasets
of Table~\ref{tab_lin_eph} with the quadratic
ephemeris (Eq.~\ref{Eq_quad_ephem}), obtaining a very coherent fit.
Although the depths of the minima differ, of course,
for each passband, the phase of the primary mid-eclipse
is the same for all of them.

Note that \citet{Ondrich2003} derived a smaller value of the period
decrease, $\dot P = (-2.49\pm0.18)\times10^{-9}$ days per day. For a fully
conservative mass transfer, they estimated
$\dot M = 1.4\times10^{-9}$~\ms\ per year. We note, however, that their value
refers to a mass transfer rate per {\sl one day}, and that the correct mass
transfer rate per one year should be $0.52~\mu$\ms y$^{-1}$, i.e. typical of
{\sl strongly interacting binaries} \citep{Plavec1980}. Our new result implies a mass transfer
rate that is even higher than the value by \citet{Ondrich2003} (see below).

\subsection{Spectroscopic mass ratio}
\label{sect_spec_q}
The mass ratio of the binary systems with a Roche-lobe filling component
can be estimated from the (plausible) assumption that the rotation of this
star is synchronized with the orbital revolution
\citep{Andersen1989, Harmanec1990}.
The essence of the method lies in the relative dimensions
of the Roche lobe depending solely on the binary mass ratio $q$, while the absolute
radius of the spin-orbit synchronized Roche-lobe filling star is given by
the equatorial rotational velocity, the inclination of the rotational axis
(assumed to be the same as the orbital inclination) and by the rotational
period, which is identical to the binary orbital period. The third Kepler law
is used to derive the binary separation.

To apply a similar procedure
to \uxe, we measured the rotational velocity of the secondary, $v_{2}\sin i$,
using a mean spectrum constructed from three spectra closest to the phase
of the primary mid-eclipse (phase $\phi$ = 0) and compared it with
synthetic spectra. For that, a grid of
model atmospheres was constructed using the {\sl LLmodels2} program \citep{shul2004},
which allows us to calculate 1-D plane-parallel, LTE model
atmospheres for stars with effective temperatures of 5000 K and higher.
Line absorption is taken into account directly for each line taken from
a line list selected by the user. For the spectrum synthesis, we used
the program {\sl SynthV} \citep{tsym96}.

\begin{figure}
\resizebox{\hsize}{!}{\includegraphics{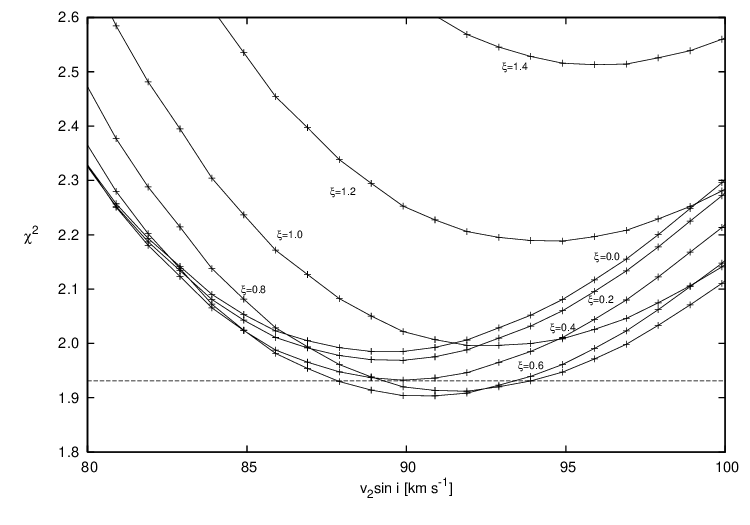}}
\caption{$\chi^{2}$ value as a function
of $v_{2}\sin i$ for various values of $\xi_{t}$. A 95\% confidence level is
shown with a dashed line.}
\label{vsini_chi2}
\end{figure}

To check whether the assumption of spin-orbit synchronism
is in agreement with the value of $q=0.8$ obtained from photometry alone
\citep{Olson1995,Olson2009},
we adopted the values derived by the same authors:
the temperature of the secondary
$T_{2} = 5500$ K and its surface gravity acceleration $\log g$ = 2.9.
Free parameters of the fit were microturbulent velocity, $\xi_{t}$,
and $v_{2}\sin i$. We also allowed for the adjustment of
the local continuum of the observed spectrum to that of the synthetic
one, compensating for possible large-scale misalignments due to an imperfect
continuum placement during the spectrum reduction.
To avoid the broad Balmer and stronger telluric lines, the analysis was restricted
to the 4897 to 5777~\AA\ wavelength range. The best fit was obtained for
$\xi_{t} = (0.68 + 0.23)$ km s$^{-1}$ and
$v_{2}\sin i = 90.7 \pm 3.0$ km s$^{-1}$. The $\chi^{2}$ values as a function
of $v_{2}\sin i$ for various values of $\xi_{t}$ are shown in Fig.~\ref{vsini_chi2}.

We used the value of $v_{2}\sin i = 90.7 \pm 3.0$ km s$^{-1}$ and
the orbital period $P = 5.9044365$ days for the reference epoch derived in
the previous section (first term in Eq. \ref{Eq_quad_ephem}). In addition, we used
the inclination $i = 89.5^{0}$ and the RV semi-amplitude of the
orbital velocity of the secondary $K_{2} = 108.3$ km s$^{-1}$ derived
by \citet{Olson1995}. Following the numerical procedure presented
in the appendix of \citet{Harmanec1990} (see also the beginning of this section),
we arrived at $q = 1.3$.
This estimate of $q$ differs substantially from the value of $q=0.8$
derived by \citet{Olson1995}.
Note that we estimate the mass ratio again with the same method but
using new values of all relevant parameters derived in this paper
at the beginning of Sect.\ref{Sect_Disc}.

\begin{figure}
\resizebox{\hsize}{!}{\includegraphics{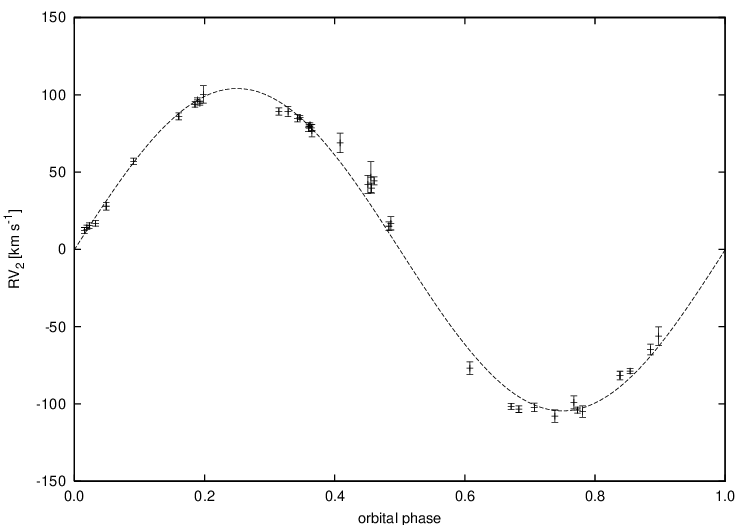}}
\caption{The RV curve of the secondary given by Eq.~(\ref{EqR2Curve})
is shown by a dashed sinusoidal curve. The individual mean RVs of the
secondary, based on SPEFO measurements, with their error bars are also shown.}
\label{RVCurves}
\end{figure}

To find a more accurate value of the mass ratio, $q$,
we used all of the spectra at our disposal (cf. Sect.~\ref{SectSpect}).
The RVs of individual unblended spectral lines were measured with SPEFO, which
displays direct and reverse traces of the line
profiles superimposed on the computer screen and the user can slide
them to achieve their exact overlapping for the studied detail of the
profile.
The orbital phases were calculated with the quadratic
ephemeris (Eq.~\ref{Eq_quad_ephem}). Many suitable strong lines of the secondary
could be identified. The RV curve of the secondary was
calculated by performing a non-linear least squares fit assuming a circular
orbit
\begin{equation}
RV_{2}=(104.3\pm1.1)\sin(2\pi\phi)\ \mathrm{km\ s}^{-1}-0.2\pm0.8\ \mathrm{km\ s}^{-1}.
\label{EqR2Curve}
\end{equation}
The resulting theoretical RV curve is plotted with the individual
RV measurements in Fig. \ref{RVCurves}.
A systematic deviation of the measured RVs from the calculated ones is visible
before the secondary eclipse ($\phi=0.5$). We attribute this difference to the Rossiter-–McLaughlin effect.
We return to this question in later sections.

\begin{figure}
\resizebox{\hsize}{!}{\includegraphics{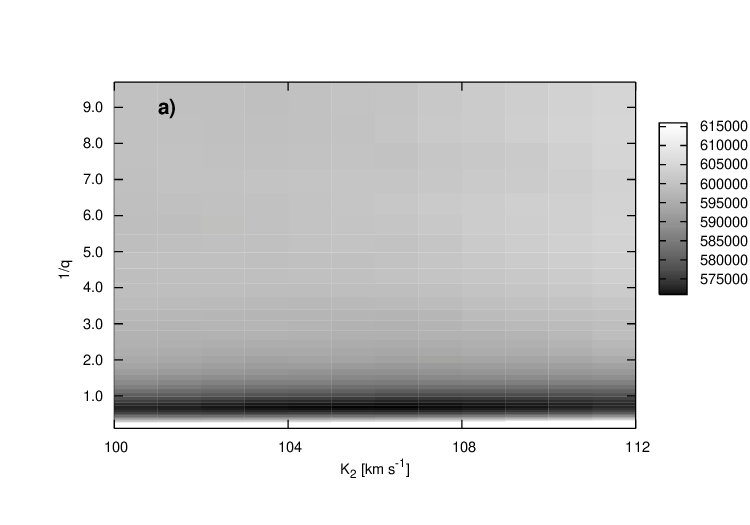}}
\resizebox{\hsize}{!}{\includegraphics{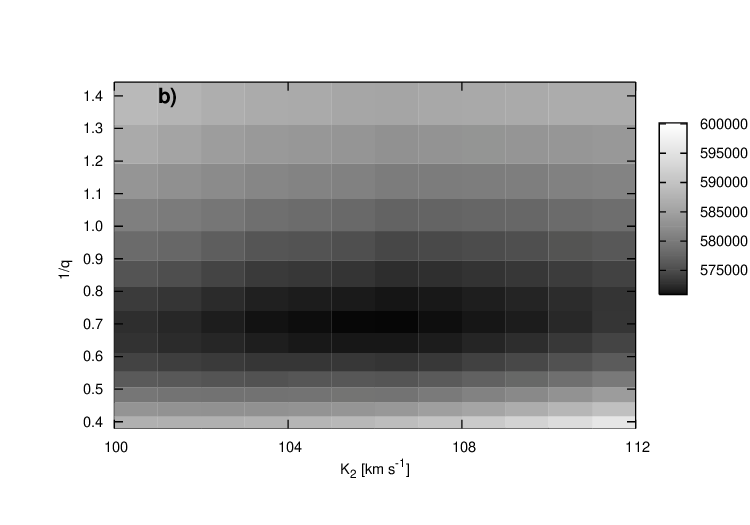}}
\caption{{\bf a)} Sum of squares of residuals, $ssr$, calculated with KOREL as a function of converged values of
radial velocity semi-amplitude of the secondary, $K_{2}$, and inverse mass ratio
$1/q$ shown as a map plot. {\bf b)} The same as in {\bf a)} but with $1/q$ axis scaled for better clarity.}
\label{ak1}
\end{figure}

To find spectral lines originating in the atmosphere of the primary
proved to be a much more difficult task. A comparison of the spectra obtained
near phases $\phi_{1} = 0$ and $\phi_{2} = 0.5$ revealed that the most
promising candidates are the following iron lines:
\ion{Fe}{II}~4923.921~\AA, \ion{Fe}{II}~5018.434~\AA, and
\ion{Fe}{II}~5169.030~\AA. However, the direct RV measurements with SPEFO,
even for spectra from which the secondary lines with proper velocity
shift were subtracted, resulted in a very uncertain RV curve displaying a large
scatter. We, therefore, attempted the spectral disentangling
using the latest publicly available version of the program KOREL
\citep{Hadrava2004a}.
We concentrated on finding the best values of two parameters:
the RV semi-amplitude
of the secondary, $K_{2}$, and the mass ratio, $q$.
We derived many solutions mapping the plausible range of both parameters.
Since the value of $K_{2}$ is fairly well constrained by the
solution given in Eq.~(\ref{EqR2Curve}) and by an independent result by \citet{Olson1995}
($K_{2}=108.3$~\ks), we only inspected the $K_2$ in the range from 100
to 112~\ks. On the other hand, we treated the mass ratio as unconstrained
over a wide range of values between 0.1 and 10.
We allowed convergence only for the time of the primary epoch and kept all
other parameters fixed.

In Fig.~\ref{ak1}, we plotted the sum of squares of the residuals, $ssr$,
as a function of the two parameters $K_{2}$ and $q$ as a map plot.
Although the Fig.~\ref{ak1} represents only a crude mapping of the parameter space, it is
quite clear (see Fig.~\ref{ak1}a)) that the best-fit solution lies in the region  $0.5 <1/q < 1.0$.

To obtain more accurate values of the parameters $K_{2}$ and $q$,
we selected 20 initial parameter sets with the lowest value of $ssr$ and
allowed for the convergence of $K_{2}$ and $q$ while keeping everything else fixed.
We obtained $K_{2}= 104.6$~\ks and $1/q = 0.87$ as the values of the lowest $ssr$.
By estimating the error in $1/q$ to be 0.05 and
taking the error in $K_{2}$ from
Eq. \ref{EqR2Curve}, we end up with
\begin{equation}
\begin{array}{ccl}
K_{2}&=&104.6\pm1.1\mathrm{\ km\ s}^{-1},\\
q&=&1.15\pm0.05,\\
K_{1}&=&120.6\pm5.4\mathrm{\ km\ s}^{-1},\\
M_{1}\sin^{3} i&=& 3.24\pm 0.38\mathrm{\ }M_{\odot},\\
M_{2}\sin^{3} i&=& 3.74\pm 0.28\mathrm{\ }M_{\odot},\\
a\sin i &=& 26.29\pm 0.64\mathrm{\ }R_{\odot}.
\end{array}
\label{Eq_RVs_and_q}
\end{equation}

\begin{figure}
\begin{center}
\resizebox{\hsize}{!}{\includegraphics{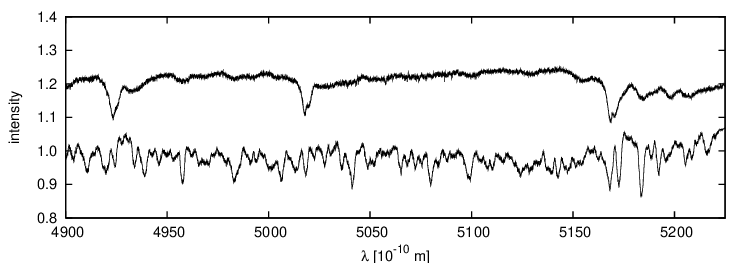}}
\caption{The disentangled spectra of the two components.}
\label{DisFeLines}
\end{center}
\end{figure}
Fig. \ref{DisFeLines} shows the disentangled spectra of both components. The spectrum of the primary
is shifted upwards by 0.2 for clarity.
The three strong Fe {\footnotesize II} lines are clearly visible in the spectrum of the primary.

In the approximation of fully conservative mass transfer, a period decrease in
time leads to the mass ratio $q>1$. \citet{Ondrich2003} carried a similar analysis
and came to the same conclusion, challenging the value of $q=0.8$ published
by \citet{Olson1995}.  Our new value of $q=1.15$ agrees with
their analysis.

Using the well-known formula for fully conservative mass transfer (cf e.g. \citet{Harmanec1993}) and our
estimates of the masses from the RV measurements, we obtain
a mass transfer rate of $\dot{M}= (-3.6\pm 1.3)$ $\mu {\rm M}_{\odot}y^{-1}$.
This value for the mass transfer rate is typical of W Ser stars
\citep{Plavec1980, Plavec1992}.

\begin{figure}
\begin{center}
\resizebox{\hsize}{!}{\includegraphics{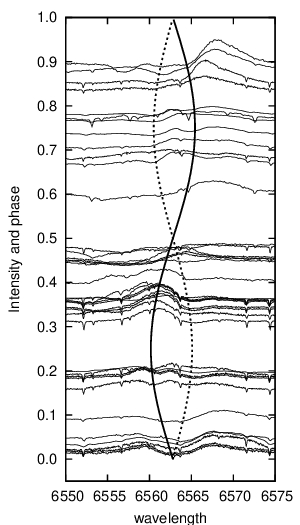}}
\caption{A plot of all 39 H$\alpha$ profiles available to us, arranged
according to their orbital phases (via appropriate shift of their continua
along the ordinate labeled with the orbital phases). The solid
and dotted sinusoids represent the orbital motion of the primary
and secondary, respectively.}
\label{halpha2d}
\end{center}
\end{figure}

We also note that the region around H$\alpha$ exhibits a strong
emission at all phases (Fig.~\ref{halpha2d}).
All 39 available spectra are shown. One can see that changes in the H$\alpha$
profile are linked to orbital motion of the primary and secondary, shown
in Fig.~\ref{halpha2d} by solid and dotted sinusoids, respectively.

\subsection{Light curve modeling}

Considering that mass transfer is present in the system and that \citet{Ondrich2003}
found the emission lines of Si {\footnotesize IV} around 1400 {\AA} and of
C {\footnotesize IV} around 1550 {\AA} in the UX Mon spectra, we felt
encouraged to model the system as a \object{W Ser} star.
This required us to model an optically and geometrically thick disk
surrounding the primary star.

We used the BINSYN suite of programs \citep{Linnell1984, Linnell1996, Linnell2000} with a
parameter optimization procedure based on the steepest descent method developed
by DS. The optimization procedure controlled the flow of BINSYN software,
automating the process.

Using the results from the previous section, we kept $q=1.15$ and
$M_{1}\sin^{3}i = 3.24\textrm{ }M_{\odot}$ constant. As can be seen from the light curve,
the eclipse is total or close to totality.
Therefore,  we can estimate the mass of the primary to be in the range
of 3.24 $M_{\odot}$ to 3.4 $M_{\odot}$. Assuming that the primary is a main-sequence
star, we estimate its radius $R_{1} \approx 3\textrm{ }R_{\odot}$ and
temperature $T_{1} \approx 13000$ K \citep{Harmanec1988}. Note that this assumption
probably underestimates the actual radius because such large stars in close binaries
are probably inflated with respect to their main-sequence counterparts.

Although the effect of the mass transfer can easily be seen from the secularly
changing period, its effect on the physical parameters of the system
(e.g. masses and separation between the components) is below the
accuracy of light curve modeling.
Therefore, we can assume that these values have been
constant over the past fifty years covered by the observations.
Hence, it should be legitimate to fold all the observations into a single
phase diagram using the quadratic ephemeris (Eq.~\ref{Eq_quad_ephem}).
We used the value of the period $P_{0}$, given in Table \ref{FOTELeph}, for the reference epoch
as the input to BINSYN.

\begin{table}
\caption{Initial and final values of the system parameters obtained with BINSYN. Errors of first four listed parameters were taken
from Table~\ref{FOTELeph} and Eq.~\ref{Eq_RVs_and_q}.}
\label{TabDisk1}
\begin{center}
\begin{tabular}{c c c c}
\hline\hline
parameter & initial value & converged value & error\\
\hline
$P_{0}$ [days]& 5.9044365 & -- & 0.0000020\\
$M_{1}\sin^{3}i$ [$M_{\odot}$] & 3.24 & -- & 0.38 \\
$a\sin i$ [$R_{\odot}$] & 26.29 & -- & 0.64\\
$q$ & 1.15 & -- & 0.05 (est.)\\
$e$ & 0.0 & -- & -- \\
$i$ [$^{0}$] & 80.50 & 80.51 & 1.0  (est.)\\
$\Omega_{1}$ & 8.77 & -- & 0.10 (est.) \\
$\Omega_{2}$ & 4.00 & 3.98 & 0.02 (est.) \\
$F_{1}$ & 1.0 & -- & -- \\
$F_{2}$ & 1.0 & -- & -- \\
$T_{1}$(pole) [K] & 13000 & -- & 1000 (est.)\\
$T_{2}$(pole) [K] & 5000 & 5989 & 200 (est.)\\
$\dot{M}$ [$\mu M_{\odot}$y$^{-1}$]& 3.60  & 5.46 & 0.50  (est.)\\
$R_{A}$ [$R_{\odot}$] & 9.2  & 9.0 & 1.0 (est.)\\
$H_{V}$ [$R_{\odot}$] & 3.5 & 2.9 & 0.2 (est.)\\
\hline
\end{tabular}
\end{center}
\end{table}

We combined the datasets (4), (7), (8), (9), and (11) from Table \ref{PhotData} into three groups
corresponding to the
Johnson's $UBV$ filters and used the steepest descent method together with BINSYN to obtain the best
fit of the system parameters. The solution was found by fitting the parameters
of the system simultaneously in all three passbands.

\begin{figure}
\begin{center}
\resizebox{\hsize}{!}{\includegraphics{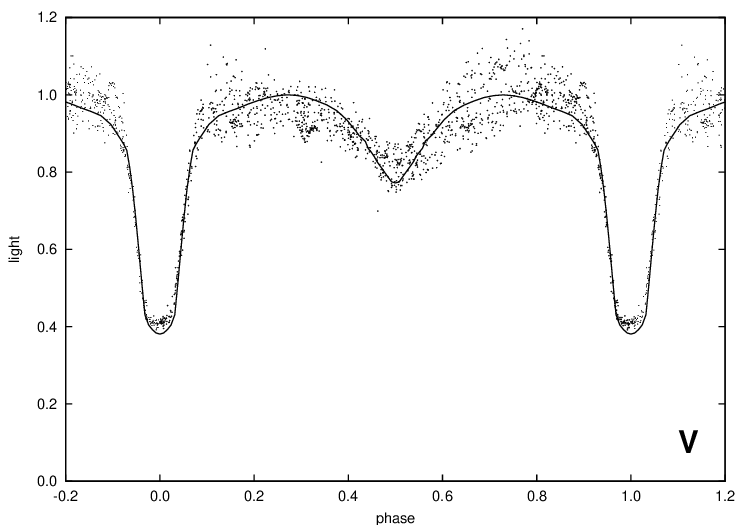}}
\resizebox{\hsize}{!}{\includegraphics{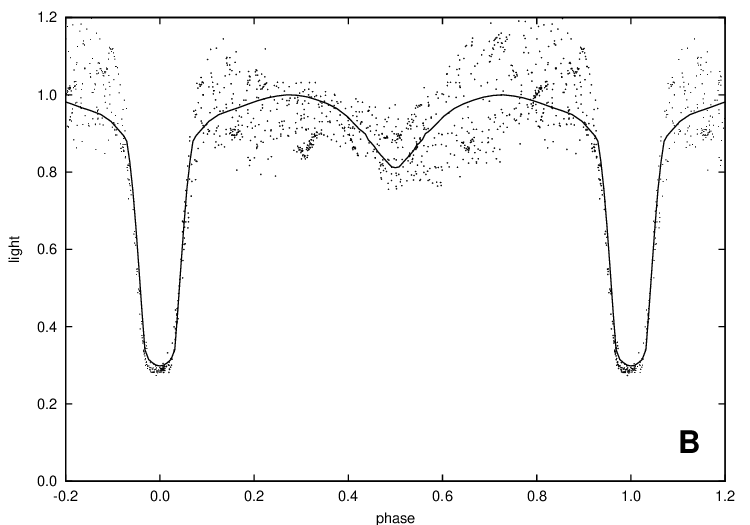}}
\resizebox{\hsize}{!}{\includegraphics{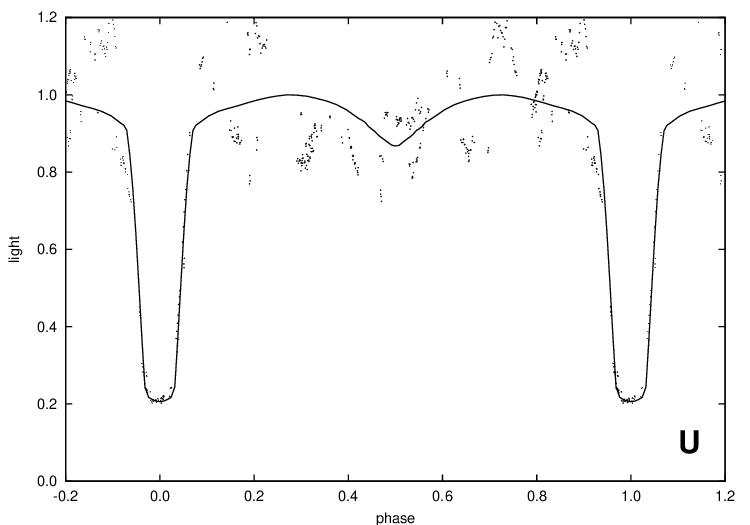}}
\caption{Observed data and light curves calculated for the final set of parameters (Table \ref{TabDisk1}).}
\label{LCSolDisk1}
\end{center}
\end{figure}

\begin{figure}
\begin{center}
\resizebox{\hsize}{!}{\includegraphics{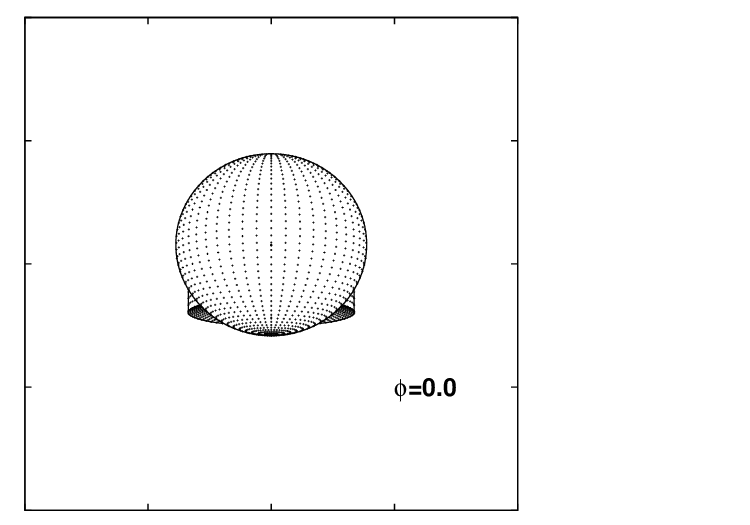}}
\resizebox{\hsize}{!}{\includegraphics{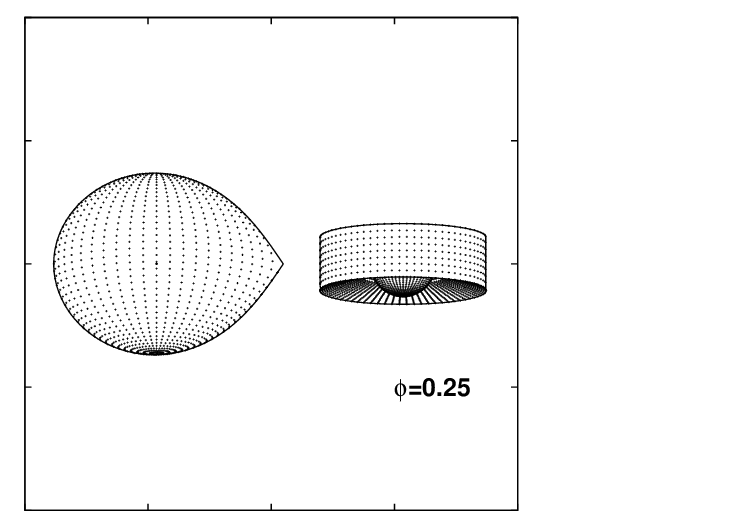}}
\resizebox{\hsize}{!}{\includegraphics{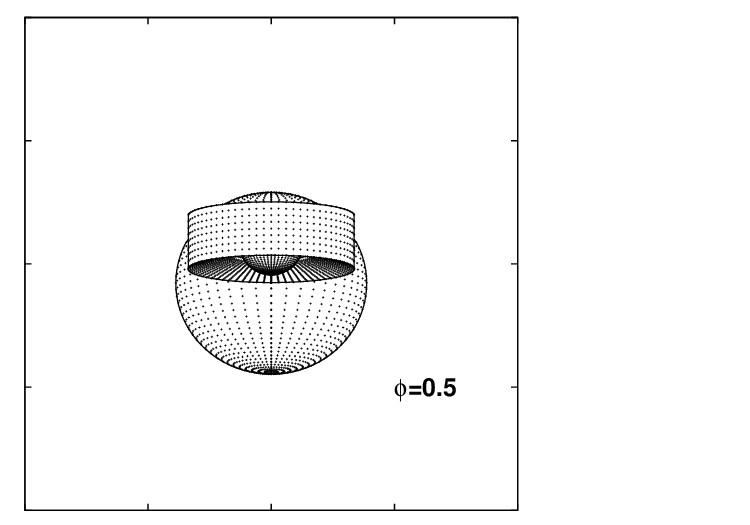}}
\caption{Three views of the binary system UX Mon with an optically thick disk around the primary.}
\label{RealViewDisk1}
\end{center}
\end{figure}

Table \ref{TabDisk1} lists initial and final values of the system parameters.
For the albedo and gravity brightening coefficients, we used the standard
values for a radiative envelope of the primary and a convective envelope of the secondary.
The initial values of photospheric potentials ($\Omega_{1}$ and
$\Omega_{2}$) and inclination, were chosen by trial and error.
The same is true for the outer radius and height of the disk ($R_{A}$
and $H_{V}$). The inner radius of the disk, $R_{B}$, was set to be equal
to that of the star.

Since we expect the secondary to be an evolved star that almost fills its Roche lobe,
we assume synchronized rotation ($F_{2}=1$).
The situation is quite the opposite when we consider the primary. Because it might gain momentum
from the accreting material, we expect it to rotate faster. However, initial
trials showed that this parameter has no visible effect on the resulting light curve.
This is beacuse the primary is almost invisible through the accretion disk
surrounding it. To reflect this, we set the formal value of this parameter $F_{1}=1$, but its final
value is completely uncertain. Our model does not offer the possibility to measure this
parameter more accurately.

\begin{table}
\caption{Parameters of the UX Mon derived from the values listed in
Table \ref{TabDisk1}.}
\label{TabFinalSolution}
\begin{center}
\begin{tabular}{l c}
\hline\hline
parameter & value \\
\hline
$M_{1}$ [$M_{\odot}$] & 3.38$\pm$0.40 \\
$M_{2}$ [$M_{\odot}$] & 3.90$\pm$0.29 \\
$R_{1}$(pole) [$R_{\odot}$] & 3.49$\pm$0.05 \\
$R_{2}$(pole) [$R_{\odot}$] & 9.80$\pm$0.03 \\
$a$ [$R_{\odot}$]& 26.65$\pm$0.65 \\
$\log g_{1}$(pole)  & 3.88$\pm$ 0.01 \\
$\log g_{2}$(pole)  & 3.061$\pm$0.003\\

$T_{D}$(rim) [K]& 6245$\pm$200 \\
$T_{D}$(face) [K]& 7487$\pm$200 \\
$l_{1}$ & 0.39$\pm$0.03 \\
$l_{2}$ & 0.44$\pm$0.01 \\
$l_{D}$(rim) & 0.14$\pm$0.03 \\
$l_{D}$(face) & 0.03$\pm$0.01 \\

\hline
\end{tabular}
\end{center}
\end{table}

The calculated and observed light curves in three
passbands are shown in Fig. \ref{LCSolDisk1}.
In Fig. \ref{RealViewDisk1}, we show three views
of UX Mon at different orbital phases. The shape of the large secondary corresponds to its Roche geometry
close to the critical, Roche-lobe filling case.
The primary star, which is mainly eclipsed by the disk, is outlined for enhanced clarity.

\begin{figure}
\begin{center}
\resizebox{\hsize}{!}{\includegraphics{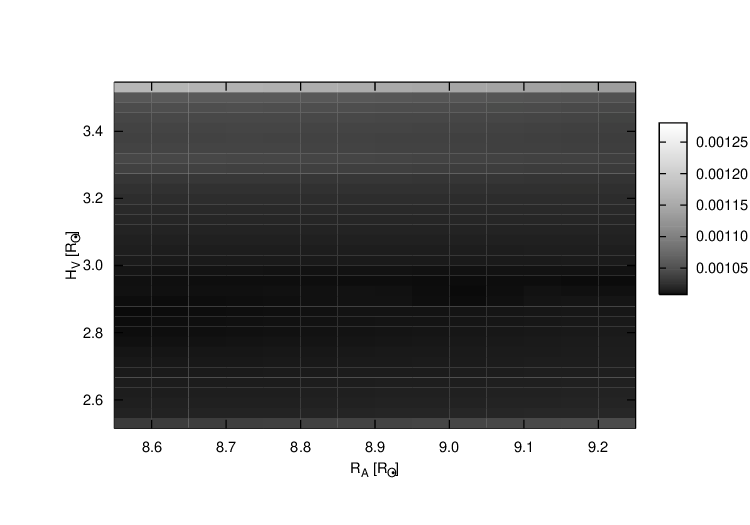}}
\caption{A map plot of $rms$ vs. $R_{A}$ and $H_{V}$ in the vicinity of the light curve solution.}
\label{map_RA_HV}
\end{center}
\end{figure}

We also investigated the uncertainty in the parameters describing
the geometry of the disk, $R_{A}$ and $H_{V}$. The quality of the model
fit for the various values of both parameters in the neighborhood
of the best-fit solution is shown in Fig.~\ref{map_RA_HV}. One can see that
the uncertainty in the height of the disk $H_{V}$ is much smaller
than that of the outer disk radius $R_{A}$. Our estimates of the errors
in both parameters, given in Table~\ref{TabDisk1}, are based on the
mapping presented in Fig.~\ref{map_RA_HV}.

In Table \ref{TabFinalSolution}, we give some important parameters of UX Mon derived
from the values shown in Table \ref{TabDisk1}. Note that, in both tables, certain errors are not indicated
as estimated values, although they are calculated via a parameter whose error {\it is} estimated (mainly
through connection with inclination, $i$, and mass ratio, $q$).
Our estimate of the error in the inclination, $i$, is based on the work
of \citet{Linnell2006}, who also used the BINSYN software package in their analysis.
We admit, however, that \ux is a more complicated system because of the
circumstellar matter around the primary. Consequently, the true
uncertainty in the determination of $i$ could even be several degrees.
The estimate of the error in the mass ratio, $q$ = 0.05, is subjective,
based on our experience with the sensitivity of KOREL in finding the mass
ratio (Sect~\ref{sect_spec_q}). We also admit that the parameter mapping
shown in Fig.~\ref{ak1} leaves us with some uncertainty regarding the error
in $q$, which should not, however, be larger than $\pm$0.1.
The values of $T_{D}$ are the temperatures of the disk and $l_{1}$, $l_{2}$, $l_{D}$(rim), and $l_{D}$(face) are luminosities relative
to the system luminosity of the primary, secondary, rim of the disk, and face
of the disk, respectively.
Luminosities are calculated at phase $\phi=0.25$ for a wavelength of
$\lambda=$ 550 nm.

\section{Discussion}
\label{Sect_Disc}

Using the stellar parameters obtained from our light curve solution, we can repeat the procedure of
estimating the mass ratio, $q$, described at the beginning of section \ref{sect_spec_q}.
By first using, $T_{2}$(back) = 5742 K and $\log g_{2}$(back) = 2.83, we get $\xi_{t} = (0.5\pm0.25)$ km s$^{-1}$ and
$v_{2}\sin i = 83.0\pm0.8$ km s$^{-1}$. The minimum $\chi^{2}$ obtained for this model is $\sim$1.5, i.e. $\sim$1.25 times smaller than
the $\chi^{2}$ obtained from the model described at the beginning of section \ref{sect_spec_q}.

\begin{figure}
\begin{center}
\resizebox{\hsize}{!}{\includegraphics{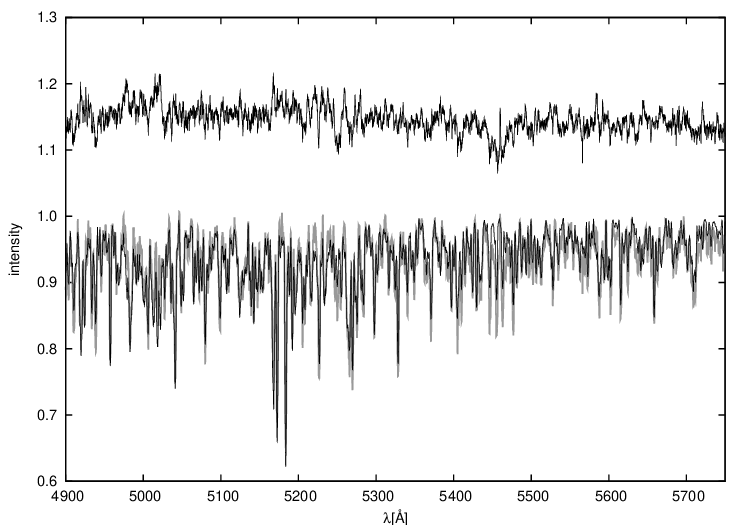}}
\caption{Observed (gray) and calculated spectra (black, superimposed over the observed spectrum) for
$T_{2}$(back) = 5742 K and $\log g_{2}$(back) = 2.83. The difference spectrum is shifted upwards and shown in black.}
\label{SpectrumFit}
\end{center}
\end{figure}

\begin{figure}
\begin{center}
\resizebox{\hsize}{!}{\includegraphics{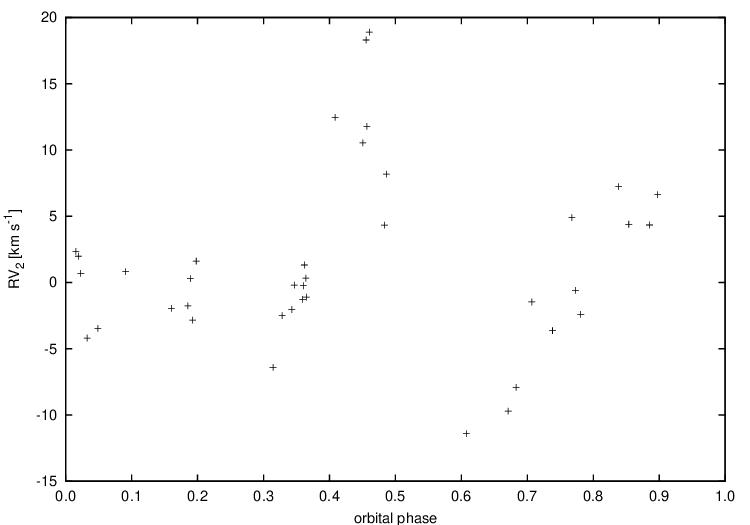}}
\caption{Difference between the measured RVs and those calculated from Eq. \ref{EqR2Curve}.}
\label{Diff_RVs}
\end{center}
\end{figure}

In Fig. \ref{SpectrumFit}, we show the observed and synthetic spectra as well as the difference between the two.
The match is not perfect, but the deviations can be attributed to the disk that
is still visible.

We can again estimate the mass ratio, but this time using all values obtained in this paper:
$K_{2}=104.6$ km s$^{-1}$, $i = 80.5^{0}$, $P_{0} = 5.9044365$ days, and $v_{2}\sin i = 83.0$ km s$^{-1}$.
This time the estimated mass ratio turns out to be $q_{est} = 1.22$. This result is close to the value of
$q = 1.15\pm0.05$ that we found from the radial velocity analysis.

\begin{figure}
\begin{center}
\resizebox{\hsize}{!}{\includegraphics{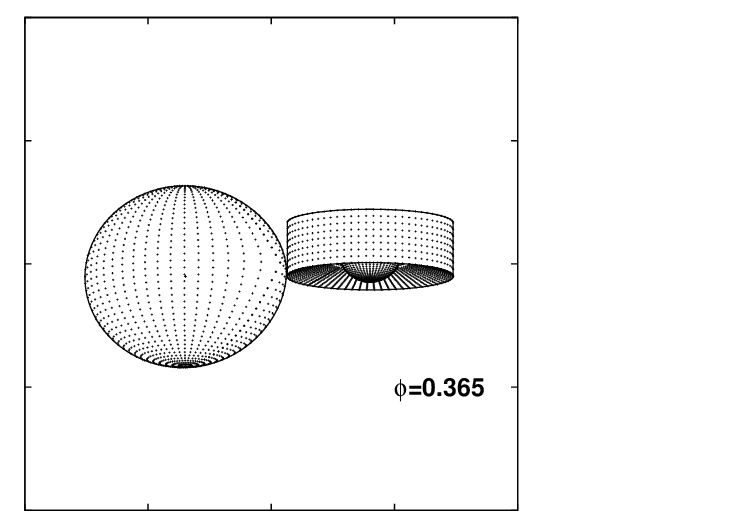}}
\caption{Phase of first contact during secondary eclipse found from photometric solution.}
\label{First_contact}
\end{center}
\end{figure}

By looking at the differences between the measured and calculated RVs, we can use the Rossiter-McLaughlin
effect to estimate the size of the secondary. We assume that the phase at which the discrepancies start ($\phi\approx$ 0.35) is the phase of
the first contact
of the secondary eclipse. The phase interval between the first contact and the center of the secondary eclipse
is then $\Delta\phi\approx$ 0.15. The distance covered by the star during this phase interval
is $K_{2}P\Delta\phi$. On the other hand, we know from geometry of the system, that
this distance is equal to $a_{2}\Delta\phi$, where $a_{2}$ is the distance of the secondary to the center of mass.
We can eliminate $a_{2}$ by using $a_{2}=R_{2}/sin(2\pi\Delta\phi)$, assuming a circular orbit and ignoring
the inclination. Finally we get
\begin{equation}
R_{2} = \frac{K_{2}P}{2\pi}\sin(2\pi\Delta\phi) \approx 9.9\mathrm{\ }R_{\odot}.
\end{equation}
This value is close to the side radius of the disk that we found from our photometric solution $R_{2}$(side) = 10.3 $R_{\odot}$.
From Fig. \ref{First_contact}, we see that the eclipse starts at $\phi\approx$ 0.365, which is a little later than
we estimated from the RV measurements alone ($\phi\approx$ 0.35).

\section{Principal results and open questions}
The following principal results of our study lead to a substantial
revision of the currently accepted model of the system:
\begin{enumerate}
\item The orbital period of the system is secularly decreasing
at a rapid rate of 0.26~s per year.
\item The analysis of spectroscopic observations as well as
the (plausible) assumption that the cooler, Roche-lobe filling
component has its rotation synchronized with the orbital revolution
both lead to the conclusion that the cooler component is the {\sl more
massive of the two}.
\item Reasonable basic physical properties of the components were
obtained via modeling the light curves on the premise that there is
an optically thick accretion disk around the hotter component, which
is eclipsed in the primary minimum. We admit, however, that the
remaining uncertainties in some of the derived model parameters are
still quite large.
\item The above findings identify \ux as a binary in a rare early stage
of the mass exchange before the mass ratio reversal.
\end{enumerate}

There are, however, still a few open questions and possible alternative
views that need to be solved and clarified using new dedicated observations
and more sophisticated modeling:
\begin{itemize}
\item There is no really satisfactory and self-consistent explanation of the
true cause of non-orbital light changes clearly seen, for instance, in
Fig.~\ref{LCSolDisk1}. These variations were very
systematically investigated by \citet{Olson2009}, who noted that the
scatter in the light curve is largest in the phase interval 0.6-0.8
from the primary minimum, i.e. around the elongation with component~1
receding from us (see their Fig~2 or Fig.~\ref{LCSolDisk1} here).
These are the phases where the gas stream between
the components can be seen projected on to the disk of the primary star. They
therefore tentatively concluded that the light variations are
related to variations in the rate of the mass transfer, possibly
induced by the magnetic activity of the cool mass-losing star.
There is, however, a very interesting plot in their Fig.~8
showing cyclic light variations at phases of totality during the primary
minima with cycle lengths of some 800-1000~days. These variations are
reminiscent of similar cyclic variations found for other
strongly interacting binaries like RX~Cas \citep[517~d;][]{kalv79},
$\beta$~Lyr \citep[282~d;][]{hec96} or AU~Mon \citep[417~d;][]{Desmet2009}.
For all these three binaries, the light variations are probably strictly
periodic and \citet{Desmet2009} argue that they are caused by
variations in the circumbinary matter. All four systems, including \uxe,
are mass-exchanging binaries but only \ux is in the initial phase before
the mass ratio reversal.
\item There is also another aspect worth considering:
\citet{hec96} showed that the bulk of the \ha emission in $\beta$~Lyr
does not originate from the optically thick disk but from bipolar jets,
oriented perpendicular to the orbital plane and originating from the region
where the gas encircling the mass-gaining star hits the original gas stream flowing from
the Roche-lobe filling component \citep{bis2000}.
Using only indirect arguments (mainly the phase offset of the RV curve of
the \ha emission), \citet{Desmet2009} speculated about the possible presence of
bipolar jets also  for AU~Mon. For the moment, this possibility
cannot be excluded even for \uxe.
\item There are several possible ways to improve our
understanding of \uxe. First, spectrointerferometry of high resolution
could show what the character and geometry of the medium responsible
for the observed \ha emission is. Continuing systematic spectral and
photometric variations could reveal possible long-term changes
in the \ha emission and its relation to the -- already
known -- long-term light changes. A hydrodynamical modeling based on
the binary parameters obtained in this study could also show whether
our tentative model is compatible with the predictions from the theory.
Finally, a more sophisticated
modeling of the light curves including the effects of
various possible circumstellar structures, appears necessary, and
not only for \uxe.
\item Considering how crucial \ux might be to achieving a clearer understanding
of the process of large-scale mass exchange in this type of system, it seems
obvious that the continuation of its systematic investigation is worth the effort.
\end{itemize}

\begin{acknowledgements}
We are very grateful to Prof. E.C.~Olson and Mr.~R.~ Meyer for providing
us with their observations. We are also grateful to
Dr. A.P.~Linnell for providing us with the BINSYN software
package.
We thank Mr.~D.~Ond\v{r}ich for making preliminary
digitalization of the published photometric
observations. We also thank Mrs.~Laura~Abrami and Mrs.~Chiara~Doz for supplying
a copy of the Scaltriti paper, not available to us.
We gratefully acknowledge the initial reduction of the Ond\v{r}ejov
CCD spectrograms by Dr.~M.~\v{S}lechta. The first two Ond\v{r}ejov
spectra were obtained for us by Drs.~M.Dov\v{c}iak and M.~Wolf.
We acknowledge the use of publicly available versions of the programs
FOTEL and KOREL written by Dr.~P.~Hadrava. A constructive criticism of the original version of this study
by the referee, Dr. Andrej Pr\v{s}a, helped to improve the presentation
and is gratefully acknowledged.
Dr.~M.~Wolf and Mr.~D.~Ond\v{r}ich obtained a few
observations of \ux at Hvar.
The research of PH and partly of HB was supported by the grants 205/06/0304
and P209/10/0715 of the Czech Science Foundation.
PH was also supported from the Research Program MSM0021620860
{\sl Physical study of objects and processes in the solar system
and in astrophysics} of the Ministry of Education of the Czech Republic.
We acknowledge the use of the electronic database from CDS Strasbourg and
electronic bibliography maintained by the NASA/ADS system.
\end{acknowledgements}

\bibliographystyle{aa}
\bibliography{14920uxmon} 

\end{document}